\documentclass[prd,twocolumn,showpacs,showkeys,preprintnumbers,amsmath,amssymb,floatfix]{revtex4}

\usepackage[utf8]{inputenc}
\usepackage{hyperref}

\usepackage{dcolumn}
\usepackage{bm}
\usepackage{graphicx}
\usepackage{subfigure}

\def\be{\begin{equation}}
\def\ee{\end{equation}}
\def\bestar{\begin{equation*}}
\def\eestar{\end{equation*}}

\begin{document}

\title{Holographic superconductor from a non-commutative inspired anti-de Sitter-Einstein-Born-Infeld black hole}

\author{Marco Maceda}
\email{mmac@xanum.uam.mx}

\author{Sergio Patiño-López}
\email{selp@xanum.uam.mx}

\affiliation{Departamento de F\'{i}sica \\ Universidad Aut\'{o}noma Metropolitana - Iztapalapa\\
                   Av. San Rafael Atlixco 186, A.P. 55-534, C.P. 09340, Mexico City, M\'exico}                   
                                  
\date{\today}

\begin{abstract}
Using a noncommutative inspired anti-de Sitter-Einstein-Born-Infeld black hole with both non-commutative mass and charge distributions, we analyse a holographic superconductor in the context of the AdS/CFT correspondence. We show that the coupling of the Born-Infeld parameter and the noncommutative perturbations provides an enhancing mechanism for higher values of the expectation value of the condensate and that noncommutativity counter-balances nonlinearity of the electrodynamics to keep the ratio $\omega_g/T_c \approx 8$.
\end{abstract}


\maketitle

\section{Introduction}
\label{intro}

The AdS/CFT correspondence~\cite{Maldacena:1997re} is one of the most fruitful proposals in the last two decades.  It is a powerful tool to understand specific aspects of physical systems from two different perspectives, one arising from quantum field theory in flat spacetime and the other from string theory in curved spacetime. This link between two different theories establishes the main property of the gauge/gravity duality, mainly the existence of a strong-weak connection, that allows the analysis of a strongly coupled quantum field theory through a weakly interacting gravitational theory. 

Nowadays, holographic duality has an increasing number of applications in different areas of physics, for example, condensed matter systems, quantum information, etc. One interesting condensed matter system that has been discussed in the literature through the AdS/CFT correspondence, and its gauge/gravity duality is holographic superconductors~\cite{Hartnoll:2008vx,Kruglov:2018xzs, Jing:2010zp,Horowitz:2010gk,Musso:2014efa}. The starting point to analyse these systems is the Lagrangian
\begin{eqnarray}
S &=& \frac 1{16} \int d^4x\, \sqrt { -g } [ R - 2\Lambda + L\left( F \right) 
\nonumber \\[4pt]
&&- |\partial_\mu \psi - iq A_\mu \psi|^2 - m^2 |\psi|^2 ], 
\end{eqnarray}
where $L(F)$ is a function of the electromagnetic tensor $F_{\mu \nu}$, $A_\mu$ is the associated 4-potential and $\psi$ is a charged complex scalar field. In the original analysis by Gubser~\cite{Gubser:2008px} using Maxwell electrodynamics, the probe limit was introduced where back-reaction contributions to the background metric due to the bulk fields are neglected; the probe limit simplifies the analysis of the holographic superconductor while keeps most of the interesting physics. Following this approach and a matching procedure for the solutions~\cite{Gregory:2009fj,Roychowdhury:2012hp,Gangopadhyay:2012am}, both Maxwell and Born-Infeld (BI) electrodynamics~\cite{Born:1934gh} have been considered in the literature~\cite{Liu:2010ka,Tallarita:2010vu,Pan:2011vi,Roychowdhury:2012vj,Gangopadhyay:2013qza,Lai:2015rva,Sheykhi:2015mxb}. 

Recently, non-commutative holographic superconductors have attracted attention since these systems allow for higher expectation values of the condensation operator~\cite{Pramanik:2015eka,Ghorai:2016qwc}. These models~\cite{Nicolini:2005vd,Nicolini:2005gy,Banerjee:2009xx} use non-commutative smeared distributions for the sources, either gravitational or electromagnetic. The distributions are of the Gaussian type, incorporating the non-commutative parameter in their definitions; they provide a mechanism to replace the point-like behaviour of the sources, a feature usually linked to the presence of singularities of the gravitational and electromagnetic field. This approach gives then regular static and stationary black hole solutions in general relativity from which certain aspects, thermodynamics for example, may be analysed~\cite{Modesto:2010rv,Gonzalez:2015mpa}.

In this paper, we consider a holographic superconductor based on a non-commutative inspired anti-de Sitter-Einstein-Born-Infeld~(AdSEBI) black hole. As it is well-known, BI electrodynamics is a generalisation of Maxwell's electrodynamics, has no birefringence and provides a regularisation of the self-energy of a point-like charge~\cite{Kerner:2001qq}; the strength of the electric field, measured by the BI parameter $b$, is finite at the location of the source. Previous analyses of holographic superconductors based on BI electrodynamics exist~\cite{Zhao:2012cn} and also including other modifications~\cite{Gregory:2009fj,Pan:2011vi}. In our treatment, we follow a non-perturbative approach on the BI parameter $b$ to calculate the expectation value of the condensation operator according to the AdS/CFT duality~\cite{Ammon:2015wua,Zaanen:2015oix}. In this way, we fully explore the relationship between this parameter and the non-commutative parameter $\theta$ and see how they affect the dependence of $\langle O_2 \rangle$ on the critical temperature of the condensate.

The structure of this paper is as follows. In Sec.~\ref{secc:2} we present the general procedure to build non-commutative inspired black holes and we focus on the AdSEBI black hole with only a mass smeared distribution in (3+1)-dimensions. The discussion there serves to emphasise the key role that a charge smeared distribution has for the classical horizon radius to get non-commutative corrections. In Sec.~\ref{secc:3} we discuss the non-commutative inspired AdSEBI black hole with both smeared distributions of mass and charge and calculate the first order non-commutative corrections to the metric. After this, in Sec.~\ref{secc:4} we consider the equations of motion for the gauge and scalar fields in the presence of BI electrodynamics and the non-commutative background. Imposing the continuity of the fields and their derivatives, we find the condensation operator and critical temperature of the holographic superconductor; we always work in the probe limit throughout this section. Then in Sec.~\ref{secc:5}, we calculate the conductivity of the holographic superconductor at its critical temperature. Finally, in the Conclusions, we summarise our results and discuss further aspects to be explored. Our units are chosen as $G = c = 1$.
\section{Non-commutative inspired AdSEBI black hole}
\label{secc:2}

First, we consider the Born-Infeld (BI) electromagnetic Lagrangian
\be
L(F) = b^2 \left( 1 - \sqrt{1+ \frac {F^{\mu\nu} F_{\mu\nu}}{2b^2}} \right),
\ee
where $b$ is the BI parameter; for large values of $b$ we recover Maxwell's electrodynamics $L(F) \sim  -F^{\mu\nu} F_{\mu\nu}/4$. Since 
\be
 \frac {\partial L(F)}{\partial g^{\mu\nu}} = - \frac 12 \frac {F_{\sigma\nu} F^\sigma{}_\mu}{\sqrt{1+ F^2/2b^2}},
\ee
it has the nice property that
\be
\frac {\partial L(F)}{\partial g^{00}} = -\frac {\partial L(F)}{\partial g^{11}},\qquad \frac {\partial L(F)}{\partial g^{22}} = \frac {\partial L(F)}{\partial g^{33}} = 0,
\label{der00der11bi}
\ee
for a purely static electric field.

Now, from the conservation law
\be
\nabla_\mu \left( \frac {F^{\mu\nu}}{\sqrt{1 + F^2/2b^2}} \right) = 0,
\ee
we obtain
\be
L(F) = b^2 \left( 1 - \sqrt{1 - E(r)^2/b^2} \right),
\ee
where 
\be
E(r) = \frac q{\sqrt{r^4 + q^2/b^2}},
\ee
is the electric field due to a point-like charge $q$ located at the origin. 

Given a generic electromagnetic Lagrangian $L(F)$, the field equations
\begin{eqnarray}
R_{\mu\nu} -\Lambda g_{\mu\nu} &=& \frac {\partial L(F)}{\partial g^{\mu\nu}} + \frac 12 g_{\mu\nu} \left( - L(F) + \frac {\partial L(F)}{\partial g^{\rho\sigma}} g^{\rho\sigma} \right) 
\nonumber \\[4pt]
&&+ 8\pi \left( \mathfrak T_{\mu\nu} - \frac 12 g_{\mu\nu} \mathfrak T^\lambda{}_{\lambda} \right),
\end{eqnarray}
where~\cite{Nicolini:2005vd,Nicolini:2008aj,Banerjee:2009xx}
\be
\mathfrak T^\mu{}_{\nu} := diag (h_1, h_1, h_3, h_3), \quad h_3 := (r^2 h_1)_{,r}/2r, 
\ee
and
\be
h_1(r) := -\rho_m(r) = -\frac {\cal M}{(4\pi\theta)^{3/2}} e^{-r^2/4\theta},
\ee
define a (3+1)-dimensional non-commutative inspired black hole solution with electric charge. The normalization of the distribution $\rho_m$ is such that
\be
\int d^3 x \, \rho_m (r) = \cal M,
\ee
where $\cal M$ is the mass of the charged source, known as the bare mass and different from the ADM mass when an electromagnetic field is present~\cite{Nicolini:2008aj,Modesto:2010rv}. In this approach, the energy-momentum tensor $\mathfrak T_{\mu\nu}$ encodes non-commutativity throughout the use of a smeared distribution of mass dependent on the non-commutative parameter $\theta$; non-commutativity is incorporated in the theory through the use of non-commutative sources.

Let us consider a static spherically symmetric spacetime in $(3+1)$-dimensions
\be
ds^2 = - e^{2\mu(r)} dt^2 + e^{2\nu(r)} dr^2 + r^2 d\Omega_{(2)}.
\ee
Then, from the previous field equations, the non-commutative inspired metric coefficients for the BI electrodynamics are such that $\mu + \nu = 0$ and~\cite{Gonzalez:2015mpa}
\begin{eqnarray}
e^{-2\nu} &=& \kappa - \frac 13 \Lambda r^2 + \frac 23 b^2 r^2 \left( 1- \sqrt{1 + \frac {{\cal Q}^2}{b^2 r^4}} \right)   
\nonumber \\[4pt]
&&+ \frac 43 \frac {q^2}r \int_r^\infty \frac {ds}{\sqrt{s^4 + {\cal Q}^2/b^2}} 
\nonumber \\[4pt]
&&
- \frac {4\cal M}{\pi^{1/2}} \frac 1r \gamma\left( \frac 32, \frac {r^2}{4\theta} \right).
\label{ncsolelectric}
\end{eqnarray}
The values of $\kappa = +1, 0, -1$  determine the existence of a spherical, planar or hyperbolic horizon respectively, and $\gamma(n,z)$ is the lower gamma function~\cite{Abramowitz:1965}
\be
\gamma(n,z) := \int_0^z dt \, t^{n-1} e^{-t}.
\ee
Using the identity~\cite{Gunasekaran:2012dq}
\be
{}_2 F_1 \left( \frac 12, n ; n+1; - z \right) = n \int_0^1 \frac {t^{n - 1} dt}{(1 + zt)^{1/2}},
\ee
where ${}_2 F_1(a, b; c; z)$ is the hypergeometric function and $n$ is a constant, we obtain the equivalent expression
\begin{eqnarray}
e^{-2\nu} &=& \kappa - \frac 13 \Lambda r^2 + \frac {2b^2 r^2}3 \left(1 - \sqrt{1 + \frac {{\cal Q}^2}{b^2 r^4}} \right) 
\nonumber \\[4pt]
&& + \frac 43 \frac {q^2}{r^2} \,{}_2F_1\left( \frac 12, \frac 14; \frac 54; - \frac {{\cal Q}^2}{b^2 r^4}\right) 
\nonumber \\[4pt]
&&
- \frac {4\cal M}{\pi^{1/2}} \frac 1r \gamma\left( \frac 32, \frac {r^2}{4\theta} \right),
\label{ncmcoeff}
\end{eqnarray}
for the metric coefficient; it defines a non-commutative inspired adS-Einstein-BI (AdSEBI) spacetime. Henceforth we set $\Lambda = - 3/L^2$ and $\kappa=0$ that corresponds to the planar solution.

\subsection{Horizon radius}

In the non-commutative AdS/CFT correspondence, the horizon radius of the non-commutative inspired black hole is needed to find an expression for the condensation operator in terms of the temperature of the condensate. For this purpose, we focus in the limit $4\theta \ll r^2$ corresponding to small non-commutative perturbations. Then, for the incomplete gamma function in the metric coefficient, we can use the asymptotic expression~\cite{Abramowitz:1965}
\be
\gamma(n,z) = \Gamma(n) - e^{-z} z^{n-1} \left( 1 + \frac {a - 1}z + \dots \right),
\label{asymplgf}
\ee
where $\Gamma(z)$ is the gamma function; we obtain thus
\be
ds^2 = - f(r) dt^2 + \frac 1{f(r)} dr^2 + r^2 d\Omega_{(2)},
\ee
where
\begin{eqnarray}
f(r) &=& - \frac {2\cal M}r + \frac {r^2}{L^2} + \frac {2\cal M}{\sqrt{\pi \theta}} e^{-r^2/4\theta} 
\nonumber \\[4pt]
&&+ \frac {2b^2 r^2}3 \left(1 - \sqrt{1 + \frac {{\cal Q}^2}{b^2 r^4}} \right) 
\nonumber \\[4pt]
&& + \frac 43 \frac {{\cal Q}^2}{r^2} \,{}_2F_1\left( \frac 12, \frac 14; \frac 54; - \frac {{\cal Q}^2}{b^2 r^4}\right).
\label{aproxncmcoeff}
\end{eqnarray}
This function $f(r)$ is split as
\be
f(r) = f^c(r) + f^{\theta}(r),
\ee
where $f^c(r)$ comprises the $\theta$-independent terms and $f^\theta (r)$ involves the $\theta$-correction terms. Due to the latter, the horizon radius of the classical EBI black hole is modified. Following~\cite{Pramanik:2015eka}, we consider first the horizon radius $r_0$ of the classical extremal EBI black hole defined by the conditions
\be
f^c (r_0) = 0, \qquad f^c_{,r} (r_0) = 0.
\label{extremalc}
\ee
Explicitly, we have~\cite{Roychowdhury:2012hp,Roychowdhury:2014cva} 
\be
r_0 = \frac {\sqrt{2b{\cal Q}} L}{[3(3 + 4b^2 L^2)]^{1/4}} \sim \frac {\sqrt{{\cal Q}L}}{3^{1/4}} + O(b^{-2}).
\ee

The modified horizon radius determined by the condition $f(r_+) = 0$ can then be written as $r_+ := r_0 + \alpha$, where the non-commutative corrections are contained in the quantity $\alpha$. We have thus to lowest order
\begin{eqnarray}
0 &=& f^c (r_0) + f^c_{,r}(r_0) \alpha + \frac 12 f^c_{,rr}(r_0) \alpha^2 
\nonumber \\[4pt]
&&+ f^\theta (r_0) + f^\theta_{,r}(r_0) \alpha + \frac 12 f^\theta_{,rr}(r_0) \alpha^2.
\end{eqnarray}
The first two terms in the above expression vanish due to Eqs.~(\ref{extremalc}); we rearrange the remaining terms as
\be
\frac 12 f^c_{,rr}(r_0) \alpha^2 = - f^\theta (r_0) - f^\theta_{,r}(r_0) \alpha - \frac 12 f^\theta_{,rr}(r_0) \alpha^2.
\ee
We now exploit the fact that $f^{\theta}(r) = g^\theta(r) e^{-r^2/4\theta}$, a relation valid in general for the non-commutative perturbations due to Eq.~(\ref{asymplgf}), to write 
\begin{eqnarray}
f^{\theta}_{,r} &=& \left( g^\theta_{,r} - \frac 1{2\theta} r g^\theta \right) e^{-r^2/4\theta} =:  G_1^\theta e^{-r^2/4\theta},
\nonumber \\[4pt]
f^{\theta}_{,rr} &=& \left( g^\theta_{,rr} - \frac 1\theta r g^\theta_{,r} - \frac 1{2\theta} g^\theta + \frac 1{4\theta^2} r^2 g^\theta \right) e^{-r^2/4\theta}
\nonumber \\[4pt]
&=:& G_2^\theta e^{-r^2/4\theta}.
\end{eqnarray}
Using these relations we arrive to the equation
\be
\frac 12 f^c_{,rr}(r_0) \alpha^2 = - [g^\theta (r_0) + G^\theta_1(r_0) \alpha + \frac 12 G^\theta_2 (r_0) \alpha^2] e^{-r_0^2/4\theta},
\ee
or equivalently
\be
e^{r_0^2/4\theta} \alpha^2 = a_0 + a_1 \alpha + a_2 \alpha^2,
\label{equ4alpha}
\ee
where
\begin{eqnarray}
a_0 &=& -2 g^\theta (r_0) / f^c_{,rr}(r_0),
\nonumber \\[4pt]
a_1 &=& -2 G^\theta_1 (r_0) / f^c_{,rr}(r_0),
\nonumber \\[4pt]
a_2 &=& - G^\theta_2 (r_0) / f^c_{,rr}(r_0).
\end{eqnarray}
Eq.~(\ref{equ4alpha}) is a quadratic equation for the perturbation $\alpha$; since we are in the limit $4\theta \ll r^2$, its solution is given by
\be
\alpha = \pm \sqrt{a_0} e^{-r_0^2/8\theta} + \frac 12 a_1 e^{-r_0^2/4\theta}.
\ee
Notice that for this expression to make sense, we need $a_0 \geq 0$. To make this point clear, consider the case where only a non-commutative contribution related to the mass is present. The metric is then given by Eq.~(\ref{aproxncmcoeff}) and we have
\begin{eqnarray}
g^\theta &=& \frac {2\cal M}{\sqrt{\pi \theta}}, \qquad G_1^\theta = - \frac {\cal M}{\sqrt{\pi \theta^3}} r, 
\nonumber \\[4pt]
G_2^\theta &=& - \frac {\cal M}{\sqrt{\pi \theta^3}} + \frac {\cal M}{2\sqrt{\pi \theta^5}} r^2.
\label{funcsg}
\end{eqnarray}
Consider now use the field equation for the classical (commutative) solution
\be
r f^c_{,r} + f^c = 3 \frac {r^2}{L^2} + 2b ( b r^2 - \sqrt{{\cal Q}^2 + b^2 r^4}).  
\label{fequ}
\ee
Notice that for the extremal black hole, defined by the conditions $f^c(r_0) = 0 = f^c_{,r} (r_0)$, the field equation becomes
\be
0 = 3 \frac {r_0^2}{L^2} + 2b^2 r_0^2 - 2b \sqrt{{\cal Q}^2 + b^2 r_0^4},
\ee
implying $b > 0$ for this condition to hold; even though the BI Lagrangian at the onset is invariant under the change $b \to -b$, the extremal case does distinguish between positive and negative values of $b$. 

By taking the derivative of Eq.~(\ref{fequ}) with respect to $r$, we also obtain 
\be
2 f_{,r}^c + r f_{,rr}^c = 6 \frac r{L^2} + 4b^2 r \left( 1 - \frac {b r^2}{\sqrt{{\cal Q}^2 + b^2 r^4}} \right).
\label{sderfc}
\ee
Eqs.~(\ref{fequ}) and~(\ref{sderfc}) combine to give
\be
f_{,rr}^c = \frac 2{r^2} f^c + \frac {4b {\cal Q}^2}{r^2\sqrt{{\cal Q}^2 + b^2 r^4}}.
\label{2nderfc}
\ee
Hence, at the horizon $r_0$ of the extremal black hole we simply have
\be
f_{,rr}^c (r_0) = \frac {4b {\cal Q}^2}{r_0^2\sqrt{{\cal Q}^2 + b^2 r_0^4}}.
\label{2nderfcb}
\ee
Therefore, from Eqs.~(\ref{funcsg}) and~(\ref{2nderfc}) we get 
\begin{eqnarray}
a_0 &=& - \frac {{\cal M} r_0^2}{\sqrt{\pi \theta}} \frac {\sqrt{{\cal Q}^2 + b^2 r_0^4}}{b q^2},
\nonumber \\[4pt]
a_1 &=& \frac {{\cal M} r_0^3}{\sqrt{\pi \theta^3}} \frac {\sqrt{{\cal Q}^2 + b^2 r_0^4}}{2b q^2},
\end{eqnarray}
where
\begin{eqnarray}
0 &=& - \frac {2{\cal M}}{r_0} + \frac {r_0^2}{L^2} + \frac {2b^2 r_0^2}3 \left(1 - \sqrt{1 + \frac {{\cal Q}^2}{b^2 r_0^4}} \right) 
\nonumber \\[4pt]
&& + \frac 43 \frac {{\cal Q}^2}{r_0^2} {}_2F_1\left( \frac 12, \frac 14; \frac 54; - \frac {{\cal Q}^2}{b^2 r_0^4}\right).
\label{fc}
\end{eqnarray}
From these results we notice that $a_0 < 0$ since $b > 0$ for the extremal case as we remarked before; in consequence there are no small non-commutative corrections to the horizon radius of the extremal black hole if only a non-commutative distribution of mass is present. At this point it is worth comparing this analysis with the model discussed in~\cite{Pramanik:2015eka}, in particular with their Eq.~(7). As seen from it, the value of $a_0$ has three contributions: two positives and one negative. The positive terms arise because there is an additional non-commutative deformation of the electromagnetic field besides the deformation of the mass; this latter deformation is responsible for a negative contribution to the value of $a_0$ as we have just illustrated. With this in mind, in the next section we consider the case of a non-commutative inspired black hole defined by non-commutative distributions of mass and charge.

\section{Non-commutative inspired charged AdSEBI spacetime}
\label{secc:3}

We now proceed to discuss the non-commutative inspired AdSEBI spacetime with both non-commutative mass and charge distributions. First, we solve the conservation equations
\be
\nabla_\mu \left( \frac {F^{\mu\nu}}{\sqrt{1 + F^2/2b^2}} \right) = 4\pi \mathfrak J^\nu,
\ee
where we define
\be
\mathfrak J^\nu = [ \rho_{\cal Q} (r), \vec 0] :=  \left[ \frac {\cal Q}{(4\pi\theta)^{3/2}} e^{-r^2/4\theta}, \vec 0 \right],
\ee
as the non-commutative 4-vector current; the point-like charge has thus been replaced by a smooth distribution involving the non-commutative parameter $\theta$. The electric field is given then by
\be
E (r) = \frac {H(r)}{\sqrt{r^4 + H(r)^2/b^2}},
\ee
where 
\begin{eqnarray}
H(r) &:=& 4\pi \frac {\cal Q}{(4\pi\theta)^{3/2}} \int_0^r dz \, z^2 e^{-z^2/4\theta} 
\nonumber \\[4pt]
&=& \frac {2\cal Q}{\pi^{1/2}} \gamma\left( \frac 32, \frac {r^2}{4\theta} \right).
\end{eqnarray}
In the limit $4\theta \ll r^2$, we have
\be
H(r) = {\cal Q} \left[ 1 - 2\frac r{\sqrt{4\pi\theta}} e^{-r^2/4\theta} \right] + \dots
\ee

Having determined the electromagnetic part, the gravitational equations can be written down and solved; the gravitational field in this case is described by the following metric~\cite{Gonzalez:2015mpa} 
\begin{eqnarray}
e^{-2\nu} &=& \kappa  - \frac {4 \cal M}{\pi^{1/2}} \frac 1r \gamma\left( \frac 32, \frac {r^2}{4\theta} \right)
\nonumber \\[4pt]
&&+ \frac {r^2}{L^2} + \frac 23 b^2 r^2 \left( 1- \sqrt{1 + \frac {H^2(r)}{b^2 r^4}} \right) 
\nonumber \\[4pt]
&&+ \frac 23 \frac 1r \int_r^\infty ds \frac { - (4\pi)^2 H(s) s^3 \rho_q(s) + 2 H^2(s)}{\sqrt{s^4 + H^2(s)/b^2}}.
\label{lapseb}
\end{eqnarray}
We set as before $\kappa = 0$ corresponding to the planar solution. If we now use the ADM mass
\be
M : = \int_\Sigma (T^0{}_0 |_{matt} + T^0{}_0 |_{elect}) d\sigma,
\ee
the previous expression is written as
\begin{eqnarray}
e^{-2\nu} &=& - \frac {4 M}{\pi^{1/2}} \frac 1r \gamma\left( \frac 32, \frac {r^2}{4\theta} \right)
\nonumber \\[4pt]
&&+ \frac {r^2}{L^2} + \frac 23 b^2 r^2 \left( 1- \sqrt{1 + \frac {H^2(r)}{b^2 r^4}} \right) 
\nonumber \\[4pt]
&&+ \frac 23 \frac 1r \int_r^\infty ds \, \frac { - (4\pi)^2 H(s) s^3 \rho_q(s) + 2 H^2(s)}{\sqrt{s^4 + H^2(s)/b^2}}
\nonumber \\[4pt]
&&+2b^2 \left[ 1 - \frac 2{\sqrt{\pi}} \gamma \left( \frac 32, \frac {r^2}{4\theta} \right)  \right] J,
\label{lapseb2}
\end{eqnarray}
where
\be
J := \int_0^\infty ds \, (s^2 - \sqrt{s^4 + H(s)^2/b^2}).
\ee

In the limit $4\theta \ll r^2$, the last term in Eq.~(\ref{lapseb2}) is calculated using 
\be
1 - \frac 2{\sqrt{\pi}} \gamma \left( \frac 32, \frac {r^2}{4\theta} \right) \sim \frac r{\sqrt{\pi\theta}} e^{-r^2/4\theta},
\ee
together with the value of $J$ obtained from replacing $H(r)$ by $\cal Q$
\begin{eqnarray}
J_{H\to Q} &=& -\frac 23 \frac {{\cal Q}^2}{b^2} \int_0^\infty ds \, \frac 1{\sqrt{s^4 + {{\cal Q}^2}/b^2}}
\nonumber \\[4pt]
&=&  -\frac 23 \frac {{\cal Q}^2}{b^2} \times \Big| \frac b{\cal Q} \Big|^{1/2} \frac {4 \Gamma(\frac 54)^2}{\sqrt{\pi}}
\nonumber \\[4pt]
&=& - \frac 83 \frac {\Gamma(\frac 54)^2}{\sqrt{\pi}} \Big| \frac {{\cal Q}}b \Big|^{3/2}.
\end{eqnarray}

Small non-commutative corrections to the metric also have contributions of the following integrals
\be
I_1 := \int_r^\infty ds \, \frac {H(s) s^3 \rho_q(s)}{\sqrt{s^4 + H^2(s)/b^2}},
\ee
\be
I_2 := \int_r^\infty ds \, \frac {H^2(s)}{\sqrt{s^4 + H^2(s)/b^2}}.
\ee
The expressions for $I_1$ and $I_2$ in the limit $4\theta \ll r^2$ are in Appendix~\ref{appa}. We have then
\begin{eqnarray}
e^{-2\nu} &=& - \frac {2M}r + \frac {r^2}{L^2} + \frac 23 b^2 r^2 \left( 1- \sqrt{1 + \frac {{\cal Q}^2}{b^2 r^4}} \right) 
\nonumber \\[4pt]
&&+  \frac 43 \frac {{\cal Q}^2}{r^2} \,{}_2 F_1 \left( \frac 12, \frac 14; \frac 54; - \frac {{\cal Q}^2}{b^2 r^4} \right) + g^\theta (r) e^{-r^2/4\theta},
\label{ncmcebi}
\end{eqnarray}
where
\begin{eqnarray}
g^\theta(r) &=&  \frac {2M}{\sqrt{\pi \theta}} + \frac 23  \frac {{\cal Q}^2}{\sqrt{\pi\theta}} \frac r{\sqrt{r^4 + {\cal Q}^2/b^2}}
\nonumber \\[4pt]
&&-(4\pi)^2 \frac1{3r} \frac {{\cal Q}^2}{(4\pi\theta)^{3/2}} \frac {r^2}{\sqrt{r^4 + {\cal Q}^2/b^2}} 
\nonumber \\[4pt]
&&+ \frac 83 \frac {{\cal Q}^2}r  \sqrt{\frac \theta\pi} \frac 1{\sqrt{r^4 + {\cal Q}^2/b^2}}  \left\{ -2 + \frac {{\cal Q}^2}{b^2} \frac 1{r^4 + {\cal Q}^2/b^2} \right\}
\nonumber \\[4pt]
&&- \frac {16}3 \frac {\Gamma(\frac 54)^2}{\sqrt{\pi}} \Big| \frac {{\cal Q}}b \Big|^{3/2} \frac {b^2}{\sqrt{\pi \theta}}.
\end{eqnarray}
The non-commutative inspired AdSEBI metric is thus
\be
ds^2 = - e^{-2\nu} dt^2 + e^{2\nu} dr^2 + r^2 d\Omega_{(2)}.
\ee
From this result, we recover the well known expression for commutative EBI spacetime in the limit $\theta \to 0$. Furthermore, it is straightforward to calculate the horizon radius $r_+$ since $g^\theta(r)$ is known.

\section{Holographic $s$-wave superconductor}
\label{secc:4}

The starting point for our analysis of the holographic superconductor is the Lagrangian 
\begin{eqnarray} 
\mathcal{ L }_{ M } & = & { r }^{ 2 }{ b }^{ 2 }\left( 1-\sqrt { 1+\frac { { F }_{ \mu \nu  }{ F }^{ \mu \nu  } }{ 2{ b }^{ 2 } }  }  \right) \nonumber \\  &  & -{ r }^{ 2 }{ \left| { \partial  }_{ \mu  }\psi - iq{ A }_{ \mu  }\psi  \right|  }^{ 2 }-{ r }^{ 2 }{ m }^{ 2 }{ \left| \psi  \right|  }^{ 2 }. 
\end{eqnarray}
Here $A_\mu := (\phi, \vec 0)$ is the 4-vector potential, $\phi$ is the electrostatic potential and $F_{\mu\nu}$ is the electromagnetic tensor defined in the usual way; it follows that $F_{\mu\nu} F^{\mu\nu} = -2 \phi_{,r}^2$. Without loss of generality, the complex scalar field $\psi$ is chosen to be real. Therefore the Lagrangian becomes
\begin{eqnarray}
\mathcal{ L }_M &=& -r^2 \left( f \psi_{,r}^2 - \frac {q^2}f \phi^2 \psi^2 \right) - m^2 r^2 \psi^2 
\nonumber \\[4pt]
&&+ r^2 b^2 \left( 1 - \sqrt{1- \frac {\phi_{,r}^2}{b^2}} \right),
\end{eqnarray} 
where $f := e^{-2\nu}$ is the metric coefficient of the non-commutative inspired AdSEBI black hole. Even though $f$ contains contributions from both the non-commutative mass and charge distributions as seen from Eq.~(\ref{ncmcebi}), later we consider the probe limit where we neglect backreaction effects on the metric; this restriction means that in the background metric, we set ${\cal Q} = 0$. Nevertheless, contributions due to the charge appear indirectly since the analysis of the scalar and gauge fields involve a series development at the vicinity of the non-commutative horizon radius $r_+$.  

From the previous Lagrangian, the equation of motion for the gauge field $\phi$ is
\begin{equation}
\phi_{,rr}+\frac { 2 }{ r } \phi_{,r} \left( 1-\frac { { \phi_{,r} }^{ 2 } }{ { b }^{ 2 } }  \right) - \frac { 2{ q }^{ 2 }{ \psi  }^{ 2 } }f \phi { \left( 1-\frac { { \phi_{,r}}^{ 2 } }{ { b }^{ 2 } }  \right)  }^{ 3/2 }=0,
\end{equation}   
and for the matter field we have
\begin{equation}
\psi_{,rr} + \left( \frac { f_{,r}  }f  +\frac { 2 }{ r }  \right) \psi_{,r} - \frac { { m }^{ 2 } }f \psi + \frac {q^2 { \phi  }^{ 2 } }{ f^2   } \psi =0.
\label{comppsi1}
\end{equation}
The asymptotic behavior of the scalar and gauge fields as $r\rightarrow \infty$, where $1/f \sim  L^2/r^2$, is
\begin{equation}
\phi \left( r \right) =\mu-\frac { { \rho } }{ r } ,
\label{asinphi12}
\end{equation}
with $\mu$ and $\rho$ constants, and 
\begin{equation}
\psi =\frac { { \psi  }^{ - } }{ r } +\frac { { \psi  }^{ + } }{ { r }^{ 2 } }.
\label{asinpsi12}
\end{equation}
In this last expression we set $\psi^- = 0$, i.e. there is no source in the dual theory (standard quantisation). On the contrary, $J^+ := \psi^+$ is proportional to the expectation value of the condensation operator of the holographic superconductor using the AdS/CFT dictionary~\cite{Zaanen:2015oix}.

It is customary at this point to make the change of variable $z: = r_+/r$ to obtain a simpler expression for the equations of motion of the fields~\cite{Gubser:2008px,Pramanik:2015eka}; with $q = 1$, which is equivalent to make a rescaling of the fields to work in the probe limit, we have then the equations of motion
\begin{eqnarray}
&&\phi^{\prime\prime} + \frac { 2 z^3 }{ b^2 r_+^2 } \phi^{\prime\,3} - \frac { 2{ \psi  }^{ 2 } }f \frac {r_+^2}{z^4} \phi { \left( 1-\frac { z^4{ \phi}^{\prime\, 2 } }{ { b }^{ 2 } r_+^2}  \right)  }^{ 3/2 }=0,
\nonumber \\[4pt]
&&\psi^{\prime\prime} + \frac {f^\prime}f \psi^\prime - \frac {m^2 r_+^2}{z^4 f} \psi + \frac {r_+^2}{z^4} \frac {\phi^2}{f^2} \psi= 0.
\end{eqnarray}   
where a prime means derivative with respect to $z$. In the limit $b\to \infty$, we recover the equations of motion for the non-commutative inspired Reissner-Nordström scenario previously discussed in the literature~\cite{Pramanik:2015eka}.

Let us consider the expansion in serie of Taylor of the fields $\phi \left( z \right), $ $\psi \left( z \right) $ and the metric coefficient $f \left( z \right) $ around $z = 1$; we have
\begin{eqnarray}
 \phi \left( z \right)  & = & \phi \left( 1 \right) -\phi '\left( 1 \right) \left( 1-z \right) +\frac { 1 }{ 2 } \phi ''\left( 1 \right) { \left( 1-z \right)  }^{ 2 }+\cdots , \nonumber \\ \psi \left( z \right)  & = & \psi \left( 1 \right) -\psi '\left( 1 \right) \left( 1-z \right) +\frac { 1 }{ 2 } \psi ''\left( 1 \right) { \left( 1-z \right)  }^{ 2 }+\cdots , \nonumber \\ f\left( z \right)  & = & f\left( 1 \right) -f'\left( 1 \right) \left( 1-z \right) +\frac { 1 }{ 2 } f''\left( 1 \right) { \left( 1-z \right)  }^{ 2 }+\cdots.
\label{desafphi}
\end{eqnarray}
The series developments for the scalar and gauge fields satisfy the equations of motion if
\be
\phi \left( 1 \right) =0, \qquad \psi^\prime \left( 1 \right) =\frac { { m }^{ 2 }{ r }_{ + }^{ 2 } }{ f'\left( 1 \right)  } \psi \left( 1 \right).
\label{eq:whole}
\ee
together with
\begin{eqnarray}
&&\phi^{\prime\prime}\left( 1 \right) = - \frac { 2\phi^\prime { \left( 1 \right)  }^{ 3 } }{ { b }^{ 2 }{ r }_{ + }^{ 2 } } + \frac { 2{ r }_{ + }^{ 2 }\psi { \left( 1 \right)  }^{ 2 }\phi '\left( 1 \right)  }{ f'\left( 1 \right)  } { \left[ 1-\frac { \phi^\prime { \left( 1 \right)  }^{ 2 } }{ { b }^{ 2 }{ r }_{ + }^{ 2 } }  \right]  }^{ 3/2 },
\nonumber \\[4pt]
&&\psi^{\prime\prime}(1) = \frac 1{4f^\prime(1)^2} \left\{ [-f^{\prime\prime}(1) f^\prime(1) + 2 m^2 r_+^2 f^\prime(1) ] \psi^\prime(1) \right.
\nonumber \\[4pt]
&&\left. - [2r_+^2 \phi^\prime(1)^2 + m^2 r_+^2 f^{\prime\prime}(1) + 8 m^2 r_+^2 f^\prime(1)] \psi(1) \right\}.
\end{eqnarray}

In consequence, we have for the gauge field
\begin{eqnarray} 
\phi \left( z \right)  & = & -\phi '\left( 1 \right) \left( 1-z \right) +  \left\{ - \frac { \phi '{ \left( 1 \right)  }^{ 3 } }{ { b }^{ 2 }{ r }_{ + }^{ 2 } } \right. 
\nonumber \\[4pt]  
&  & \left. + \frac { { r }_{ + }^{ 2 }\psi { \left( 1 \right)  }^{ 2 }\phi '\left( 1 \right)  }{ f'\left( 1 \right)  } { \left[ 1-\frac { \phi '{ \left( 1 \right)  }^{ 2 } }{ { b }^{ 2 }{ r }_{ + }^{ 2 } }  \right]  }^{ 3/2 }   \right\} { \left( 1-z \right)  }^{ 2 } + \dots, 
\nonumber \\ 
& &
\label{solphi12}
\end{eqnarray}
and for the scalar field 
\begin{eqnarray} 
\psi \left( z \right)  & = & \psi \left( 1 \right) - \frac { { m }^{ 2 }{ r }_{ + }^{ 2 } }{ f'\left( 1 \right)  } \psi \left( 1 \right) \left( 1-z \right) +\frac { 1 }{ 2 } \left[ \frac { { m }^{ 4 }{ r }_{ + }^{ 4 } }{ 2f'{ \left( 1 \right)  }^{ 2 } }  \right. 
\nonumber \\[4pt]  
&  & \left. -\frac { { m }^{ 2 }{ r }_{ + }^{ 2 }f''\left( 1 \right)  }{ 2f'{ \left( 1 \right)  }^{ 2 } } -\frac { { r }_{ + }^{ 2 }\phi '{ \left( 1 \right)  }^{ 2 } }{ 2f'{ \left( 1 \right)  }^{ 2 } } - \frac {2m^2 r_+^2}{f^\prime(1)} \right] 
\nonumber \\[4pt]
&&\times \psi \left( 1 \right) { \left( 1-z \right)  }^{ 2 } + \dots. 
\label{solpsi12}
\end{eqnarray}
Here we used the relation between $\psi^\prime(1)$ and $\psi(1)$.

Following the procedure in~\cite{Pramanik:2015eka}, we match the asymptotic behavior of the fields $\phi \left( z \right) $ y $\psi \left( z \right)$ given in Eqs.~(\ref{asinphi12}) and~(\ref{asinpsi12}) with the series developments in Eqs.~(\ref{solphi12}) and~(\ref{solpsi12}) at a certain point $z=z_{0}$. We have then 
\begin{eqnarray} 
&&\mu -\frac { \rho { z }_{ 0 } }{ { r }_{ + } }  =  -\phi '\left( 1 \right) \left( 1-{ z }_{ 0 } \right) + \left\{ - \frac { \phi '{ \left( 1 \right)  }^{ 3 } }{ { b }^{ 2 }{ r }_{ + }^{ 2 } } \right.
\nonumber \\[4pt]  
&  & \left. + \frac { { r }_{ + }^{ 2 }\psi { \left( 1 \right)  }^{ 2 }\phi '\left( 1 \right)  }{ f'\left( 1 \right)  } { \left[ 1-\frac { \phi '{ \left( 1 \right)  }^{ 2 } }{ { b }^{ 2 }{ r }_{ + }^{ 2 } }  \right]  }^{ 3/2 }  \right\} (1 - z_0)^2, 
\label{pega1}
\end{eqnarray} 
and
\begin{eqnarray} 
&&\frac {J ^+ z_0^ 2}{r_+^2} = \psi \left( 1 \right) -\frac { { m }^{ 2 }{ r }_{ + }^{ 2 } }{ f^\prime\left( 1 \right)  } \left( 1-{ z }_{ 0 } \right) \psi \left( 1 \right) + \frac { 1 }{ 4 } \left[ \frac { { m }^{ 4 }{ r }_{ + }^{ 4 } }{ f'{ \left( 1 \right)  }^{ 2 } }  \right. 
\nonumber \\[4pt]  
& & \left. -\frac { { m }^{ 2 }{ r }_{ + }^{ 2 }f''\left( 1 \right)  }{ f'{ \left( 1 \right)  }^{ 2 } } - \frac { { r }_{ + }^{ 2 }\phi '{ \left( 1 \right)  }^{ 2 } }{ f'{ \left( 1 \right)  }^{ 2 } }  - \frac {4m^2 r_+^2}{f'(1)}  \right] \psi \left( 1 \right) { \left( 1-{ z }_{ 0 } \right)  }^{ 2 },
\label{jzcua}
\end{eqnarray}
Besides the continuity of the fields, we also impose the continuity of their first derivatives; we have then 
\begin{eqnarray} 
-\frac { \rho  }{ { r }_{ + } }  & = & \phi '\left( 1 \right) -2\left( 1-{ z }_{ 0 } \right) \nonumber \\  &  & \times \left\{ \frac { { r }_{ + }^{ 2 }\psi { \left( 1 \right)  }^{ 2 }\phi '\left( 1 \right)  }{ f'\left( 1 \right)  } { \left[ 1-\frac { \phi '{ \left( 1 \right)  }^{ 2 } }{ { b }^{ 2 }{ r }_{ + }^{ 2 } }  \right]  }^{ 3/2 }-\frac { \phi '{ \left( 1 \right)  }^{ 3 } }{ { b }^{ 2 }{ r }_{ + }^{ 2 } }  \right\} ,  \nonumber \\ & &
\label{deriphi}
\end{eqnarray}
and
\begin{eqnarray} 
\frac { 2J^+ z_0 }{r_+^2} & = & \frac { { m }^{ 2 }{ r }_{ + }^{ 2 } }{ f'\left( 1 \right)  } \psi \left( 1 \right) - \left[ \frac { { m }^{ 4 }{ r }_{ + }^{ 4 } }{ 2f'{ \left( 1 \right)  }^{ 2 } } - \frac { { m }^{ 2 }{ r }_{ + }^{ 2 }f''\left( 1 \right)  }{ 2f'{ \left( 1 \right)  }^{ 2 } } \right. 
\nonumber \\[4pt]  
&  & \left.  -\frac { { r }_{ + }^{ 2 }\phi '{ \left( 1 \right)  }^{ 2 } }{ 2f'{ \left( 1 \right)  }^{ 2 } } - \frac {2m^2 r_+^2}{f'(1)} \right] { \left( 1-{ z }_{ 0 } \right)  }\psi \left( 1 \right) . 
\label{jzcero}
\end{eqnarray}
From Eq.~(\ref{deriphi}) we immediately obtain 
\begin{eqnarray} 
\psi { \left( 1 \right)  }^{ 2 } & = & -\frac { f'\left( 1 \right) { \left( 1-\phi '{ \left( 1 \right)  }^{ 2 }/{ b }^{ 2 }{ r }_{ + }^{ 2 } \right)  }^{ -3/2 } }{ 2{ r }_{ + }^{ 2 }\left( 1-{ z }_{ 0 } \right)  } \nonumber \\  &  & \times \left( -1+\frac { \tilde { \rho  } { r }_{ + }^{ 4 }f'{ \left( 1 \right)  }^{ 2 } }{ { r }_{ + }^{ 5 }\left[ -\phi '\left( 1 \right)  \right] f'{ \left( 1 \right)  }^{ 2 } }  \right) , 
\end{eqnarray}
where   
\be
\tilde { \rho  } :=\rho +\frac { 2\phi '{ \left( 1 \right)  }^{ 3 }\left( 1-{ z }_{ 0 } \right)  }{ { b }^{ 2 }{ r }_{ + } }.
\ee
We also define the superconductor temperature 
\begin{equation}
{ T } \equiv \frac { 1 }{ 2\pi  } \sqrt { \frac { f'\left( 1 \right) ^{ 2 } }{ { r }_{ + }^2 }  },
\label{temp123}
\end{equation}
and the critical temperature
\begin{equation}
{ T }_{ c } \equiv \frac { 1 }{ 2\pi  } \sqrt { \frac { \tilde { \rho  }  f'\left( 1 \right) ^{ 2 } }{ r_+^3 \left[ -\phi '\left( 1 \right)  \right]  }  }.
\label{tcrit}
\end{equation}  
If $T\sim { T }_{ c }$, we have
\begin{equation}
\psi { \left( 1 \right)  }=\sqrt { -\frac { f'\left( 1 \right) { \left( 1-{ \phi^\prime (1) }^{ 2 }/{ b }^{ 2 }{ r }_{ + }^{ 2 } \right)  }^{ -3/2 } }{ { r }_{ + }^{ 2 }\left( 1-{ z }_{ 0 } \right)  }  } \sqrt { 1-\frac { { T } }{ { T }_{ c } }  } .
\end{equation}
As $b \to \infty$, we recover the expressions discussed in~\cite{Pramanik:2015eka} for the non-commutative Reissner-Nordström case. 

Combining now Eqs.~(\ref{jzcua}) and~(\ref{jzcero}), we obtain the expression  
\begin{equation}
{ J }^{ + }=\frac {r_+^2}{ z_0 } \left[ 1-\frac { { m }^{ 2 }{ r }_{ + }^{ 2 } }{ 2f'\left( 1 \right)  } \left( 1-{ z }_{ 0 } \right)  \right] \psi(1).
\label{jmasdef}
\end{equation}
Let us define ${\cal J}^+ := J^+/r_+^2$. The expectation value of the condensation operator is then 
\begin{equation}
\langle { O }_{ 2 } \rangle :=\sqrt { 2 } { \cal J }^{ + }{ r }_{ + }^{ 2 }.
\end{equation}
We have explicitly the expression
\begin{eqnarray} 
&&\langle { O }_{ 2 } \rangle = \frac { \sqrt { 2 } r_+^2}{ { z }_{ 0 } } \sqrt { -\frac { f^\prime{ \left( 1 \right)  }{ \left( 1-{ \phi^\prime (1) }^{ 2 }/{ b }^{ 2 }{ r }_{ + }^{ 2 } \right)  }^{ -3/2 } }{ { r }_{ + }^{ 2 } }  } 
\nonumber \\[4pt]  
&  & \times \sqrt { \frac { 1 }{ 1-{ z }_{ 0 } }  } \left[ 1-\frac { { m }^{ 2 }{ r }_{ + }^{ 2 } }{ 2f^\prime \left( 1 \right)  } \left( 1-{ z }_{ 0 } \right)  \right] \sqrt { 1-\frac { T }{ { T }_{ c } }  },
\end{eqnarray}
or equivalently
\begin{eqnarray} 
\frac {\langle { O }_{ 2 } \rangle }{T_c^2}  &=&  \frac { \sqrt { 2 } \left( 2\pi  \right) ^{ 2 } }{ { z }_{ 0 } } \frac { { r }_{ + }^4 }{ f^\prime{ \left( 1 \right)  }^{ 2 } } \sqrt { -\frac { f^\prime{ \left( 1 \right)  }{ \left( 1-{ \phi^\prime (1) }^{ 2 }/{ b }^{ 2 }{ r }_{ + }^{ 2 } \right)  }^{ -3/2 } }{ { r }_{ + }^{ 2 } }  } 
\nonumber \\[4pt]  
&  & \times \sqrt { \frac { 1 }{ 1-{ z }_{ 0 } }  } \left[ 1-\frac { { m }^{ 2 }{ r }_{ + }^{ 2 } }{ 2f^\prime\left( 1 \right)  } \left( 1-{ z }_{ 0 } \right)  \right] \sqrt { 1-\frac { T }{ { T }_{ c } }  },
\label{condop1}
\end{eqnarray}
where we have multiplied the former expression by a factor of $r_+^2$ and used Eq.~(\ref{temp123}) to introduce $T^2 \sim T_c^2$.

To evaluate Eq.~(\ref{condop1}), we need expressions for $f^\prime(1), f^{\prime\prime}(1)$ and $\phi^\prime(1)$; they are calculated in Appendix~\ref{appb}. Using Eqs.(\ref{prefac1}) and~(\ref{abc}), we finally obtain for the expectation value of the condensate
\begin{eqnarray} 
\frac { \langle { O }_{ 2 } \rangle  }{ { T }_{ c }^{ 2 } }  & = & a \Lambda \left( 1 - \frac {{ \phi^\prime}^{ 2 }}{{ b }^{ 2 }{ r }_{ + }^{ 2 }} \right)^{ -3/4 }  \sqrt { 1-\frac { T }{ { T }_{ c } }  }, 
\label{condop2}
\end{eqnarray}
where 
\begin{eqnarray} 
a &:=& \frac { \sqrt { 2 } \left( 2\pi  \right) ^{ 2 }{ L }^{ 3 } }{ 3\sqrt{3}{ z }_{ 0 } \sqrt{1 - z_0}} \left[ 1+\frac { { m }^{ 2 }{L}^{ 2 } }{ 6 } \left( 1-{ z }_{ 0 } \right) \right],
\nonumber \\[4pt]
\Lambda &:=& 1 - \frac {18 + 5 m^2 L^2 (1 - z_0)}{2[6 + m^2 L^2(1-z_0)]} (\alpha + \beta - \gamma),
\nonumber \\[4pt] 
\alpha  & := & \frac { { L }^{ 2 } }{ 3r_+^2 } g^{ \theta  }\left( r_+ \right) { e }^{ -r_+^2/4\theta  }, 
\nonumber \\[4pt] 
\beta  & := & \frac { { L }^{ 2 } }{ 3r_+ } g_{,r}^{ \theta  }\left( r_+ \right) { e }^{ -r_+^2/4\theta  }, 
\nonumber \\[4pt] 
\gamma  & := & \frac {{ L }^{ 2 } }{ 3 } \frac { 1 }{ 2\theta  } g^{ \theta  }\left( r_+ \right) { e }^{ -r_+^2/4\theta  }.
\end{eqnarray}
These variable encode the non-commutative corrections to the expectation value of the condensation operator; Eq.~(\ref{phibr}) gives the quotient $\phi^{\prime\,2}/r_+^2$. For simplicity, we set $L=1$ in the following.

\begin{figure}[h!]
\begin{center}
\includegraphics[width=8cm]{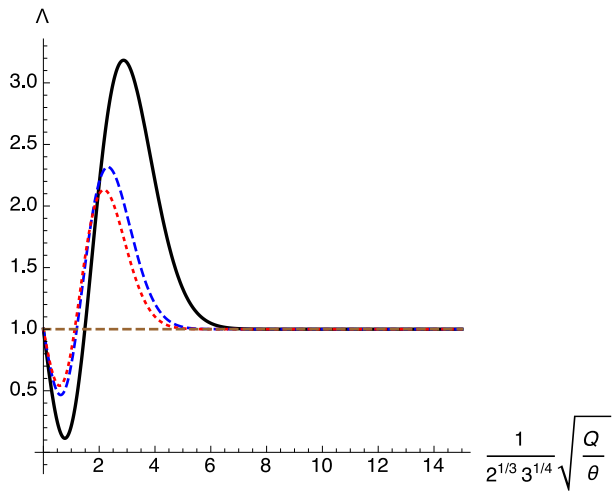}
\caption{Plot of the non-commutative factor $\Lambda$ as a function of $\lambda = 2^{-1/3} 3^{-1/4}\sqrt{{\cal Q}/\theta}$ for $b = 0.6, 1.6$, and $16$ (solid black, dashed blue and dotted red lines respectively). The horizontal dashed (brown) line corresponds to the commutative value $\Lambda = 1$; we fixed $L = 1, m^2 L^2 =-2, z_0 = 1/2$.}
\label{fig1}
\end{center}
\end{figure}

In Fig.~\ref{fig1} we plot the behaviour of the factor $\Lambda$ in terms of the variables $\lambda := 2^{-1/3} 3^{-1/4}\sqrt{{\cal Q}/\theta}$ and $b$ with $m^2 L^2 =-2, z_0 = 1/2$; $\lambda$ is the standard variable used in the non-commutative Reissner-Nordström case. As we see from the plot, the difference with respect to the constant commutative value is enhanced for small values of $b$ and as $b$ goes to infinity, we recover the behaviour associated to the  non-commutative Reissner-Nordström black hole.

\begin{figure}[h!]
\begin{center}
\includegraphics[width=8.5cm]{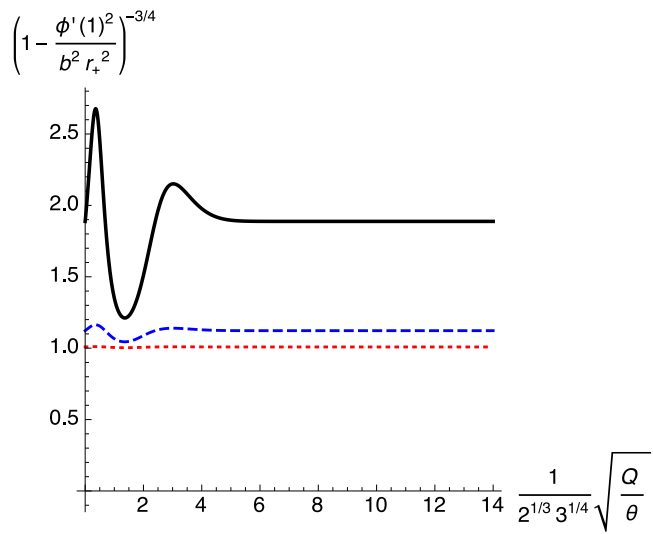}
\caption{Plot of the second factor in front of the square root in the second line of Eq.~(\ref{condop2}) as a function of $\lambda = 2^{-1/3} 3^{-1/4}\sqrt{{\cal Q}/\theta}$ for $b = 7, 14$, and $50$ (solid black, dashed blue and dotted red lines); we fixed $L= 1, m^2 L^2 =-2, z_0 = 1/2$.}
\label{fig2}
\end{center}
\end{figure}

In Fig.~\ref{fig2} we plot the second factor in front of the square root in the second line of Eq.~(\ref{condop2}) with $m^2 L^2 =-2, z_0 = 1/2$. When $\lambda =0$, the value of the second factor is $(1 - 28 b^{-2})^{-3/4}$; this is also the asymptotic value as $\lambda \to \infty$; it becomes one as $b\to\infty$ with $L$ fixed.

\begin{figure}[h!]
\begin{center}
\includegraphics[width=8.5cm]{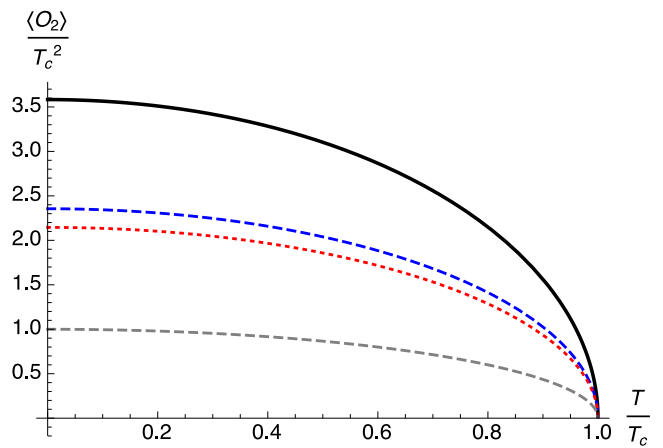}
\caption{Plots of the expectation value of the condensate for $b = 7, 14$ and $50$ (top to bottom). The lower dashed curve represents the standard commutative Reissner-Nordström case. The scale on the vertical axis is in units of the constant $a$ defined in the main text.}
\label{fig3}
\end{center}
\end{figure}

In Fig.~\ref{fig3} we show the expectation value of the condensate as a function of the ratio $T/T_c$. The (black) solid line corresponds to the values $b = 7, \Lambda_{max} = 2.17668$, the (blue) dashed line to $b = 14, \Lambda_{max} = 2.1705$ and the (red) dotted line to $b = 50, \Lambda_{max} = 2.16859$; the values of the factor $(1 - \phi^{\prime\,2}/r_+^2)^{-3/4}$ are $3.583, 2.355$ and $2.145$ respectively. The combination of the BI and the non-commutative parameters produces higher values of the condensate.

\section{Conductivity}
\label{secc:5}

Conductivity is one the main properties that characterize holographic superconductors and it is a natural consequence of the AdS/CFT correspondence~\cite{Herzog:2002fn,Herzog:2007ij,Horowitz:2008bn,Horowitz:2009ij,Hartnoll:2009sz}. This property has been extensively discussed for several configurations of black holes, including non-linear electrodynamics, higher curvature terms, Horava-Lifshitz gravity and Maxwell strength field corrections~\cite{Gregory:2009fj,Siopsis:2010uq,Pan:2011vi,Zhao:2012cn,Sheykhi:2015mxb,Mansoori:2016zbp,Sherkatghanad:2017edj,Sheykhi:2018mzs}; in the following we consider conductivity in the probe limit. Assuming the standard Ansatz $A_\mu = (\phi, 0, \delta A_x, 0 )$ where $\delta A_x := \delta A e^{-i\omega t + i k y}$ is the perturbation of the gauge potential in coordinates $(t,r, x, y)$, a straightforward calculation shows that in the case of BI electrodynamics the gauge perturbation $\delta A$ must satisfy the following differential equation
\begin{eqnarray}
f^2 \delta A_{,rr} + \left( f{f_r} + \frac 1{b^2} \frac {\phi_{,r} \phi_{,rr} f^2}{1 - \phi^2_{,r}/b^2} \right) \delta A_{,r} 
\nonumber \\[4pt]
+ \left( \omega^2 - 2 f \psi^2 (1 - \phi^2_{,r}/b^2)^{1/2} \right) \delta A = 0.
\label{conductivity}
\end{eqnarray}
To discuss the effect of nonlinearity and non-commutativity on $\delta A$, but without losing generality, we consider the situation when the superconductor is at its critical temperature $T_c$. This assumption allows on one hand, to drop out the term proportional to $\psi$ in the above differential equation; moreover, it simplifies the equation of motion for the gauge field $\phi$ in such a way that we can also obtain an explicit expression for $\phi_{,r}$ as a function of $r$. With all these ingredients, it is straightforward to solve Eq.~(\ref{conductivity}) numerically and calculate the conductivity using a few lines of code in MATHEMATICA~\cite{Hartnoll:2009sz}. 

\begin{widetext}

\begin{figure}[htbp]
\centering
\subfigure[\,$b = 0.1, \theta = 0.1$] 
{
    \label{fig:subfig1}
    \includegraphics[width=0.45\textwidth]{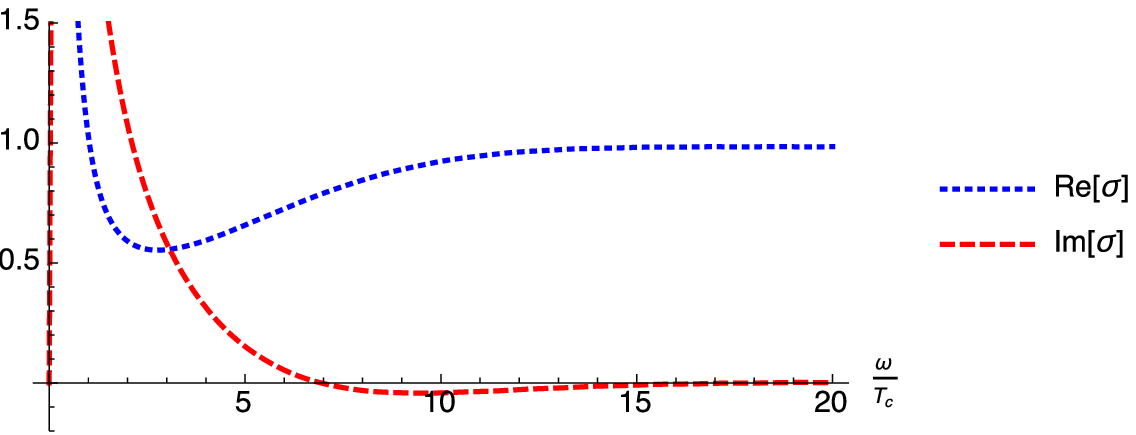}
}
\subfigure[\,$b = 0.1, \theta = 1$] 
{
    \label{fig:subfig2}
    \includegraphics[width=0.45\textwidth]{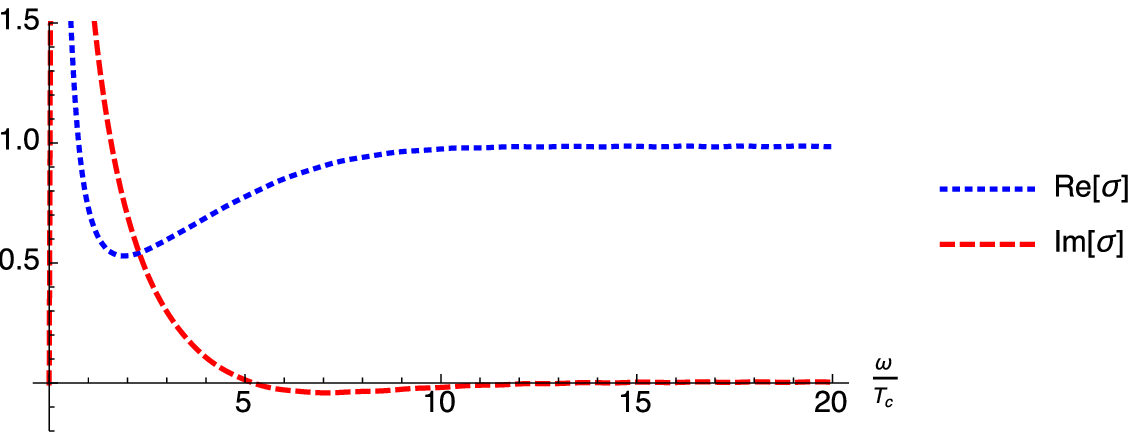}
}
\subfigure[\,$b = 1, \theta = 0.1$] 
{
    \label{fig:subfig3}
    \includegraphics[width=0.45\textwidth]{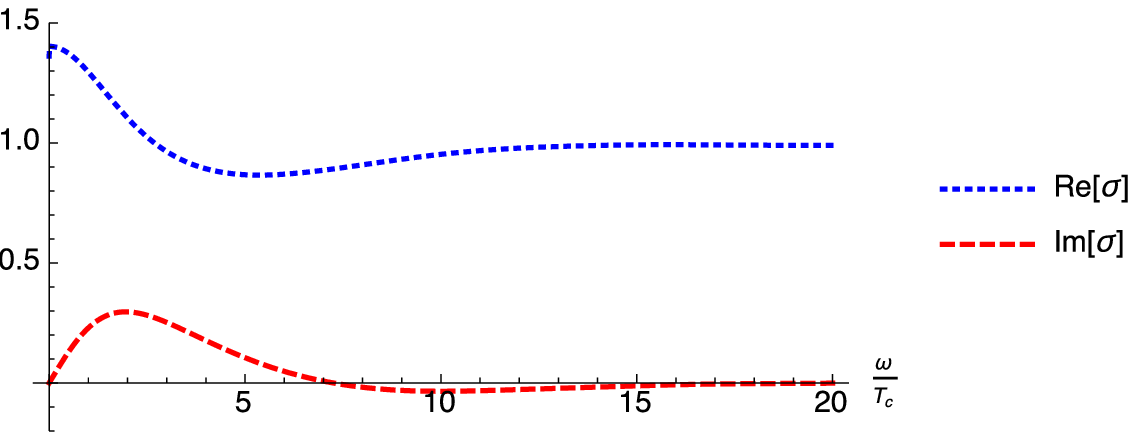}
}
\subfigure[\,$b = 1, \theta = 1$] 
{
    \label{fig:subfig4}
    \includegraphics[width=0.45\textwidth]{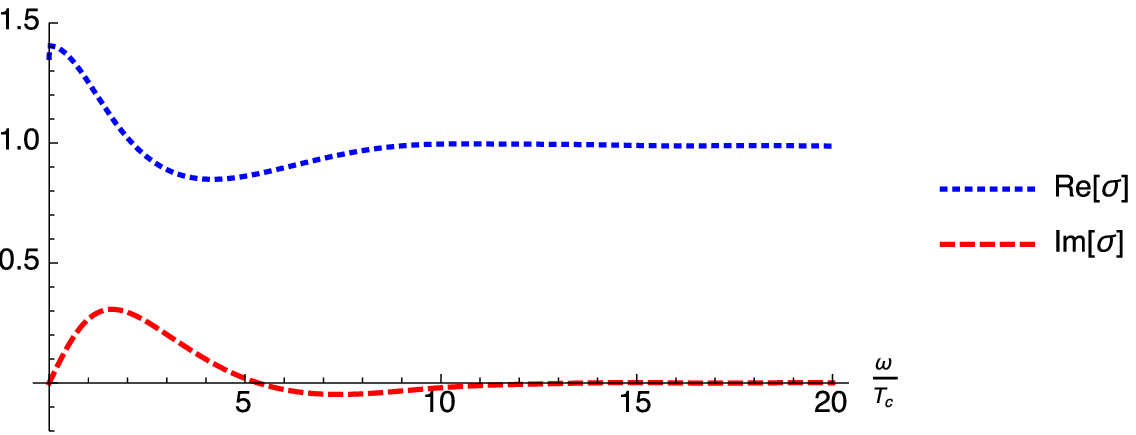}
}
\subfigure[\,$b = 10, \theta = 0.1$] 
{
    \label{fig:subfig5}
    \includegraphics[width=0.45\textwidth]{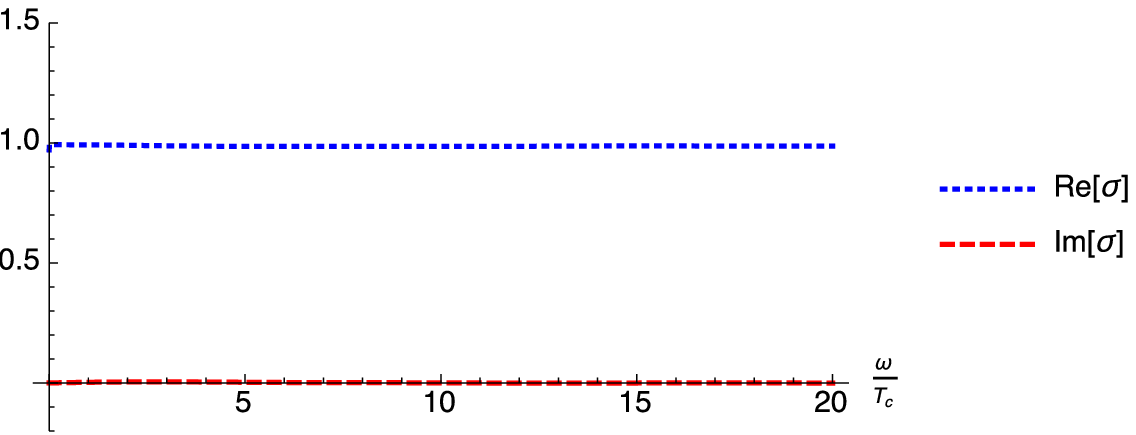}
}
\subfigure[\,$b = 10, \theta = 1$] 
{
    \label{fig:subfig6}
    \includegraphics[width=0.45\textwidth]{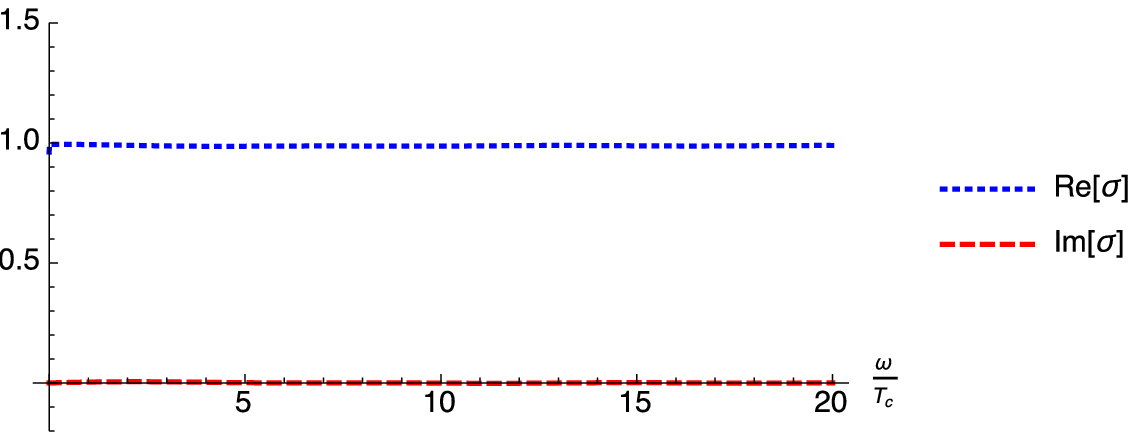}
}
\caption{Plots of the real and imaginary parts of the holographic conductivity $\sigma$ for several values of the BI parameter $b$ and the noncommutative parameter $\theta$. For a fixed value of $b$ and small noncommutativity, there are deviations from the ratio $\omega_g/T_c \approx 8$ where the minimum of $\Im \sigma$ is located; increasing the value of $\theta$ moves this minimum closer to $\omega_g/T_c \approx 8$. As $b$ increases, the conductivity becomes equal to one and the effect of noncommutativity is not noticeable.}
\label{fig:globfig} 
\end{figure}

\end{widetext}

In Fig.~\ref{fig:globfig}, we show the numerical calculation of the conductivity for different values of $b$ and $\theta$. High nonlinearity ($b$ small) together with small noncommutativity produce deviations from the ratio $\omega_g/T_c \approx 8$; this effect due to nonlinearity has been noticed before, see for example~\cite{Sheykhi:2018mzs} where similar curves were obtained. From these plots we see that when $\theta$ increases its value, the minimum of the imaginary part of the conductivity gets shifted back to a position closer to the ratio $\omega_g/T_c \approx 8$; we may say then that noncommutativity counter-balances the nonlinearity coming from the electrodynamics.

\section{Conclusions}

Using a non-commutative inspired AdSEBI black hole as a background, we analysed the behaviour of the gauge and scalar fields in the probe limit within the context of the AdS/CFT correspondence. A series development and a matching procedure for the gauge and scalar fields led to the calculation of all the relevant quantities needed to find the expectation value of the condensation operator $\langle O_2 \rangle$ according to the AdS/CFT dictionary.

Our calculations are non-perturbative on the BI parameter $b$ and all the quantities needed to find $\langle O_2 \rangle$ make use of the horizon radius $r_+$ of the non-commutative inspired black hole. The expression for the non-commutative horizon radius is somewhat cumbersome to calculate; in the particular scenario of an extremal black hole and small non-commutative modifications, we gave explicit expressions exact on $b$. The analysis of higher corrections require full knowledge of $r_+$ as a perturbative development in powers of $\exp(-r_0^2/4\theta)$.

We showed that the non-commutative modifications to the classical result for the condensate appear in two different factors. The coupling of the BI parameter $b$ with the non-commutative parameter $\theta$ produces higher values of the first factor $\Lambda$ in the expression for $\langle O_2 \rangle$ when compared with the commutative ($\Lambda = 1$) and non-commutative Reissner-Nordström case. Besides this, a second factor arises directly from the BI electromagnetic Lagrangian; it acts as a modulating amplitude becoming one in the limit $b\to\infty$ with $L$ and $\theta$ fixed. These two factors finally combine to provide an enhancing mechanism allowing higher values of the condensate. We also showed that the effect of the noncommutativity on the conductivity $\sigma$ is to counter-balance the nonlinearity induced by the electrodynamics, which shifts the ratio $\omega_g/T_c \approx 8$ where the minimum of $\Im \sigma $ is present.

The discussion of holographic superconductors should at some point include back-reaction effects, even if they are hard to analyse. Some previous treatments discuss the case of BI electrodynamics to lowest order on the BI parameter $b$~\cite{Sheykhi:2015mxb}. It would be interesting to include these effects in the presence of a non-commutative inspired gravitational background in a non-perturbative way. An approach based on the Sturm-Liouville method along the lines of~\cite{Siopsis:2010uq,Pan:2011vi,Sheykhi:2018mzs} should also be worth discussing.

\appendix
\section{Calculation of small non-commutative corrections to AdSEBI spacetime}
\label{appa}

The integrals $I_1$ and $I_2$ defined in the main text are calculated in the limit $4\theta \ll r^2$ with $H(r)$ replaced by $\cal Q$; we have thus
\be
I_1 = \frac {{\cal Q}^2}{(4\pi\theta)^{3/2}} \int_r^\infty ds \frac {s^3 e^{-s^2/4\theta}}{\sqrt{s^4 + {\cal Q}^2/b^2}}.
\ee
Let us perform the change of variable $u = s^2/4\theta$, then 
\begin{eqnarray}
I_1 &=& \frac12 \frac {{\cal Q}^2}{(4\pi\theta)^{3/2}} \int_{r^2/4\theta}^\infty du \frac {u e^{-u}}{\sqrt{u^2 + {\cal Q}^2/(4\theta b)^2}},
\nonumber \\[4pt]
&=&\frac12 \frac {{\cal Q}^2}{(4\pi\theta)^{3/2}} \int_{r^2/4\theta}^\infty du e^{-u} \left[ 1 + \left( \frac {\cal Q}{4\theta b u} \right)^2 \right]^{-1/2}
\nonumber \\[4pt]
&=&\frac12 \frac {{\cal Q}^2}{(4\pi\theta)^{3/2}} \int_{r^2/4\theta}^\infty du e^{-u} \sum_{k=0}^\infty \binom{-1/2}{k} \left( \frac {\cal Q}{4\theta b u} \right)^{2k}
\nonumber \\[4pt]
&=&\frac12 \frac {{\cal Q}^2}{(4\pi\theta)^{3/2}} \sum_{k=0}^\infty \binom{-1/2}{k} \left( \frac {\cal Q}{4\theta b} \right)^{2k} \int_{r^2/4\theta}^\infty du u^{-2k} e^{-u}
\nonumber \\[4pt]
&=&\frac12 \frac {{\cal Q}^2}{(4\pi\theta)^{3/2}} \sum_{k=0}^\infty \binom{-1/2}{k} \left( \frac {\cal Q}{4\theta b} \right)^{2k} \Gamma\left[ 1-2k, \frac {r^2}{4\theta} \right].
\end{eqnarray}
Using now the asymptotic behaviour for the upper incomplete gamma function~\cite{Abramowitz:1965} 
\be
\Gamma(n, x) \sim x^{n-1} e^{-x} \qquad x \to \infty,
\ee
we have
\begin{eqnarray}
I_1 &=&\frac12 \frac {{\cal Q}^2}{(4\pi\theta)^{3/2}} \sum_{k=0}^\infty \binom{-1/2}{k} \left( \frac {\cal Q}{4\theta b} \right)^{2k} \left( \frac {r^2}{4\theta} \right)^{-2k} e^{-r^2/4\theta}
\nonumber \\[4pt]
&=& \frac12 \frac {{\cal Q}^2 e^{-r^2/4\theta}}{(4\pi\theta)^{3/2}} \sum_{k=0}^\infty \binom{-1/2}{k} \left( \frac {\cal Q}{b r^2} \right)^{2k}
\nonumber \\[4pt]
&=& \frac12 \frac {{\cal Q}^2 e^{-r^2/4\theta}}{(4\pi\theta)^{3/2}} \left[ 1 + \left( \frac {\cal Q}{b r^2} \right)^2 \right]^{-1/2}
\nonumber \\[4pt]
&=& \frac {{\cal Q}^2}2 \frac {e^{-r^2/4\theta}}{(4\pi\theta)^{3/2}} \frac {r^2}{\sqrt{r^4 + {\cal Q}^2/b^2}}.
\end{eqnarray}

We now evaluate
\be
I_2 := \int_r^\infty ds \frac {H^2(s)}{\sqrt{s^4 + H^2(s)/b^2}}.
\ee
We have first that
\begin{eqnarray}
I_2 &=& {\cal Q}^2 \left\{ \int_r^\infty ds \frac 1{\sqrt{s^4 + {\cal Q}^2/b^2}} \right.
\nonumber \\[4pt]
&& - \frac 4{(4\pi\theta)^{1/2}} \int_r^\infty ds \frac {s e^{-s^2/4\theta}}{\sqrt{s^4 + {\cal Q}^2/b^2}}
\nonumber \\[4pt]
&& \left. + \frac {{\cal Q}^2}{b^2} \frac 2{(4\pi\theta)^{1/2}} \int_r^\infty ds \frac {s e^{-s^2/4\theta}}{(s^4 + {\cal Q}^2/b^2)^{3/2}} \right\}.
\end{eqnarray}
In the above expression, the first integral is nothing but the standard BI electromagnetic contribution to the metric. The second and third integral correspond to non-commutative corrections and can be evaluated in the same fashion as $I_1$ with the final result
\begin{eqnarray}
I_2 &=& {\cal Q}^2 \left\{ \frac 1r \,{}_2 F_1 \left( \frac 12, \frac 14; \frac 54; - \frac {{\cal Q}^2}{b^2 r^4} \right) \right.
\nonumber \\[4pt]
&& - \frac 4{(4\pi\theta)^{1/2}} \times 2\theta \frac {e^{-r^2/4\theta}}{\sqrt{r^4 + {\cal Q}^2/b^2}}
\nonumber \\[4pt]
&& \left. + \frac {{\cal Q}^2}{b^2} \frac 2{(4\pi\theta)^{1/2}} \times 2\theta \frac {e^{-r^2/4\theta}}{(r^4 + {\cal Q}^2/b^2)^{3/2}} \right\}
\nonumber \\[4pt]
&=& {\cal Q}^2 \left\{ \frac 1r \,{}_2 F_1\left( \frac 12, \frac 14; \frac 54; - \frac {{\cal Q}^2}{b^2 r^4} \right) - 4 \sqrt{\frac \theta\pi} \left( \phantom{\sqrt{\frac \theta\pi}} \right. \right.
\nonumber \\[4pt]
&&\left. \left. \frac {e^{-r^2/4\theta}}{\sqrt{r^4 + {\cal Q}^2/b^2}} + 2 \frac {{\cal Q}^2}{b^2} \frac {e^{-r^2/4\theta}}{(r^4 + {\cal Q}^2/b^2)^{3/2}} \right) \right\}.
\end{eqnarray}

\section{Calculation of $f^\prime(1), f^{\prime\prime}(1)$ and $\phi^\prime(1)$}
\label{appb}

In the probe limit, the metric function $f$ has the form
\be
f = -\frac {2M}r + \frac {r^2}{L^2} + g^\theta (r) e^{-r^2/4\theta},
\ee
or 
\be
f = -\frac {2M}{r_+}z + \frac {r_+^2}{L^2} \frac 1{z^2} + g^\theta (r) e^{-r^2/4\theta},
\ee
in terms of the variable $z : = r_+/r$. We have then
\be
f^\prime(z) = -\frac {2M}{r_+} - \frac {2r_+^2}{L^2} \frac 1{z^3} + \left[ g^\theta (r) e^{-r^2/4\theta} \right]_{,r} \left( -\frac {r_+}{z^2} \right).
\ee
Evaluating this expression at $z=1$, and using the relation
\be
-\frac {2M} {r_+} = - \frac {r_+^2}{L^2} - g^\theta (r_+) e^{-r_+^2/4\theta},
\ee
that is a consequence of $f(1) = 0$, we arrive to
\be
f^\prime(1) = -\frac {3r_+^2}{L^2} \left( 1 + \left[ \frac {L^2}{3r_+^2} g^\theta(r_+) + \frac {L^2}{3r_+} G_1^\theta(r_+) \right] e^{-r_+^2/4\theta} \right).
\ee
In this last expression, the radius $r_0$ may be used instead of the radius $r_+$ in the terms involving the Gaussian exponential since they correspond to small non-commutative corrections. A similar calculation shows that
\be
f^{\prime\prime}(1) = \frac {6r_+^2}{L^2} \left( 1 + \left[ \frac {L^2}{3r_+} G_1^\theta(r_+) + \frac {L^2}6 G_2^\theta(r_+) \right] e^{-r_+^2/4\theta} \right).
\ee

The value of $\phi^\prime(1)^2$ is obtained from Eq.~(\ref{jzcero}) when Eq.~(\ref{jmasdef}) is substituted in it. After some simplifications we have
\begin{eqnarray}
\frac {\phi^\prime(1)^2}{r_+^2} &=& \frac 1{1 - z_0} \left[ 4 \frac {f^\prime(1)^2}{r_+^4} - \frac {2m^2 f^\prime(1)}{r_+^2} (4 - 3z_0) \right.
\nonumber \\[4pt]
&&\left.+ m^4 \left( 1 - \frac {f^{\prime\prime}(1)}{m^2 r_+^2} \right)(1 - z_0) \right].
\end{eqnarray}
Using the previously deduced expressions for $f^\prime(1)$ and $f^{\prime\prime}(1)$, we obtain
\begin{eqnarray}
\frac {\phi^\prime(1)^2}{r_+^2} &=& \frac 1{1 - z_0} \frac 1{L^4} \left[ 36 H_1(r_+)^2 + 6 m^2L^2 (4 - 3z_0) H_1(r_+) \right.
\nonumber \\[4pt]
&&\left.+ (m^2 L^2)^2 \left( 1 - \frac 6{m^2 L^2} H_2(r_+) \right)(1 - z_0) \right],
\label{phibr}
\end{eqnarray}
where
\begin{eqnarray}
H_1(r_+) &:=& 1 + \left[ \frac {L^2}{3r_+^2} g^\theta(r_+) + \frac {L^2}{3r_+^2} G_1^\theta(r_+)  \right] e^{-r_+^2/4\theta},
\nonumber \\[4pt]
H_2(r_+) &:=& 1 + \left[ \frac {L^2}{3r_+} G_1^\theta(r_+) + \frac {L^2}6 G_2^\theta(r_+)  \right] e^{-r_+^2/4\theta}.
\end{eqnarray}
Here $g^\theta(r) = 2M/\sqrt{\pi\theta}, G^\theta_1(r) = -Mr/\sqrt{\pi\theta^3}, G^\theta_2(r) = -M/\sqrt{\pi\theta^3} + Mr^2/2\sqrt{\pi\theta^5}$ and $2M = r_+^3/L^2$. Furthermore, it is straightforward to see that
\begin{eqnarray}
&&\frac {r_+^4}{f^\prime(1)^2} \sqrt{-\frac {f^\prime(1)}{r_+^2}} \left[ 1-\frac { { m }^{ 2 }{ r }_{ + }^{ 2 } }{ 2f^\prime\left( 1 \right)  } \left( 1-{ z }_{ 0 } \right)  \right] = 
\nonumber \\[4pt]
&&\times \frac {L^3}{3\sqrt{3}}  \left[ 1 + \frac 16 m^2L^2 (1 - z_0) \right] 
\nonumber \\[4pt]
&&\times \left[ 1 - \frac {18 + 5 m^2 L^2 (1 - z_0)}{2(6 + m^2 L^2(1-z_0))} (\alpha + \beta - \gamma) \right],
\label{prefac1}
\end{eqnarray}
where
\begin{eqnarray} 
\alpha  & = & \frac { { L }^{ 2 } }{ 3{ r }_{ + }^{ 2 } } g^{ \theta  }\left( { r }_{ 0} \right) { e }^{ -{ r }_{ + }^{ 2 }/4\theta  }, 
\nonumber \\[4pt] 
\beta  & = & \frac { { L }^{ 2 } }{ 3{ r }_{ + } } g_{,r}^{ \theta  }\left( { r }_{ 0 } \right) { e }^{ -{ r }_{ + }^{ 2 }/4\theta  }, 
\nonumber \\[4pt] 
\gamma  & = & \frac {{ L }^{ 2 } }{ 3 } \frac { 1 }{ 2\theta  } g^{ \theta  }\left( { r }_{ + } \right) { e }^{ -{ r }_{ + }^{ 2 }/4\theta  }.
\label{abc}
\end{eqnarray}


\end{document}